\documentclass[aoas,MSNbibl,nameyear,seceqn,dvips]{arximspdf}
\usepackage{dcolumn}
\usepackage{graphicx}

\doi{10.1214/12-AOAS571} 
\volume{6}
\issue{4}
\pubyear{2012}
\firstpage{1971}
\lastpage{1997}

\makeatletter

\newcolumntype{d}[1]{D{.}{.}{#1}}

\newcommand{\dotsim}{\ \dot\sim\ }

\newcommand{\rrvert}{\vert}
\newcommand{\llvert}{\vert}

\newtheorem{theorem}{Theorem}
\newtheorem{lemma}{Lemma}

\newcommand{\dddot}[1]{\hspace*{1pt}\dot{\vphantom{\Psi}}\hspace*{1pt}
\ddot{\hspace*{-3.5pt}#1}}

\newcommand{\fdr}{\operatorname{Fdr}}
\newcommand{\tr}{\operatorname{tr}}
\newcommand{\mech}{\mathrm{mec}}
\newcommand{\vect}{\mathrm{vec}}
\newcommand{\bca}{\mathrm{BCa}}
\newcommand{\cov}{\mathrm{cov}}
\newcommand{\jeff}{\mathrm{Jeff}}
\newcommand{\rbd}{\operatorname{RBD}}
\newcommand{\poi}{\operatorname{Poi}}
\newcommand{\jack}{\mathrm{jack}}

\newcommand{\cala}{\mathcal{A}}
\newcommand{\calb}{\mathcal{B}}
\newcommand{\calf}{\mathcal{F}}
\newcommand{\n}{\mathcal{N}}
\newcommand{\real}{\mathbb{R}}

\newcommand{\bx}{\mathbf{x}}
\newcommand{\by}{\mathbf{y}}
\newcommand{\bone}{\mathbf1}

\newcommand{\hate}{\hat{E}}
\newcommand{\hatg}{\hat{G}}
\newcommand{\hatq}{\hat{Q}}
\newcommand{\hatt}{\hat{t}}
\newcommand{\hatu}{\hat{U}}
\newcommand{\hatv}{\hat{V}}
\newcommand{\halp}{\hat\alpha}
\newcommand{\hbet}{\hat\beta}
\newcommand{\bmet}{\bolds{\eta}}
\newcommand{\hbmet}{\hat{\bmet}}
\newcommand{\hgam}{\hat\gamma}
\newcommand{\hmu}{\hat\mu}
\newcommand{\bmu}{\bolds\mu}
\newcommand{\hbmu}{\hat{\bmu}}
\newcommand{\hpi}{\hat\pi}
\newcommand{\hthe}{\hat\theta}
\newcommand{\hfdr}{\widehat{\fdr}}
\newcommand{\nsig}{\Sigma}
\newcommand{\hnsig}{\hat{\nsig}}

\newcommand{\barc}{\bar{c}}
\newcommand{\barr}{\bar{r}}
\newcommand{\bars}{\bar{s}}
\newcommand{\bart}{\bar{t}}

\newcommand{\barsd}{\overline{\operatorname{sd}}}
\newcommand{\barcor}{\overline{\operatorname{cor}}}
\newcommand{\barcov}{\overline{\mathrm{cov}}}
\newcommand{\barcv}{\overline{\operatorname{cv}}}
\newcommand{\hcv}{\widehat{\mathrm{cv}}}
\newcommand{\hse}{\widehat{\mathrm{se}}}

\newcommand{\tild}{\tilde{D}}
\newcommand{\tilf}{\tilde{f}}
\newcommand{\tilr}{\tilde{R}}
\newcommand{\tilt}{\tilde{t}}
\newcommand{\tildel}{\tilde\Delta}
\newcommand{\tilpi}{\tilde\pi}
\newcommand{\tilxi}{\tilde\xi}

\newcommand{\fthe}{f_\theta}
\newcommand{\fhthe}{f_{\hthe}}

\makeatother

\begin{document}
\begin{frontmatter}

\title{Bayesian inference and the parametric bootstrap}
\runtitle{Bayesian inference and the parametric bootstrap}

\begin{aug}
\author[A]{\fnms{Bradley} \snm{Efron}\corref{}\thanksref{t1}\ead[label=e1]{brad@stat.stanford.edu}}
\runauthor{B. Efron}
\affiliation{Stanford University}
\address[A]{Stanford University\\
390 Serra Mall\\
Stanford, California 94305\\
USA} 
\end{aug}

\thankstext{t1}{Supported in part by NIH Grant 8R01 EB002784 and by
NSF Grant DMS-08-04324/12-08787.}

\received{\smonth{5} \syear{2012}}
\revised{\smonth{5} \syear{2012}}

%
\begin{abstract}
The parametric bootstrap can be used for the efficient computation of
Bayes posterior distributions. Importance sampling formulas take on an
easy form relating to the deviance in exponential families and are
particularly simple starting from Jeffreys invariant prior. Because of
the i.i.d. nature of bootstrap sampling, familiar formulas describe
the computational accuracy of the Bayes estimates. Besides
computational methods, the theory provides a connection between
Bayesian and frequentist analysis. Efficient algorithms for the
frequentist accuracy of Bayesian inferences are developed and
demonstrated in a model selection example.
\end{abstract}

%
\begin{keyword}
\kwd{Jeffreys prior}
\kwd{exponential families}
\kwd{deviance}
\kwd{generalized linear models}
\end{keyword}

\end{frontmatter}

\section{Introduction}\label{sec1}

This article concerns the use of the parametric bootstrap to carry out
Bayesian inference calculations. Two main points are made: that in the
comparatively limited set of cases where bootstrap methods apply, they
offer an efficient and computationally straightforward way to compute
posterior distributions and estimates, enjoying some advantages over
Markov chain techniques; and, more importantly, that the parametric
bootstrap helps connect Bayes and frequentist points of view.

The basic idea is simple and not unfamiliar: that the bootstrap is
useful for importance sampling computation of Bayes posterior
distributions. An important paper by \citet{newton} suggested a version
of nonparametric bootstrapping for this purpose. By ``going
parametric'' we can make the Bayes/bootstrap relationship more
transparent. This line of thought has the advantage of linking rather
than separating frequentist and Bayesian practices.

Section~\ref{sec2} introduces the main ideas in terms of an elementary
one-parame\-ter example and illustrates a connection between Jeffreys
invariant prior density and second-order accurate bootstrap confidence
limits. Both methods are carried out via reweighting of the original
``raw'' bootstrap replications. The calculation of posterior
distributions by bootstrap reweighting is a main theme here, in
constrast to Markov chain methods, which strive to directly produce
correctly distributed posterior realizations.

Multidimensional exponential families, discussed in Section \ref
{sec3}, allow the Bayes/bootstrap conversion process to be explicitly
characterized. Two important families, multivariate normal and
generalized linear models, are investigated in Sections~\ref{sec4} and
\ref{sec5}. Jeffreys-type priors can yield unsatisfactory
results in multiparameter problems [\citet{ghosh}], as shown here by
comparison with bootstrap confidence limits.

An advantage of bootstrap reweighting schemes is the straightforward
analysis of their accuracy. Section~\ref{sec6} develops accuracy
estimates for our methodology, both internal (How many bootstrap
replications are necessary?) and external (How much would the results
vary in future data sets?). The latter concerns the frequentist
analysis of Bayesian estimates, an important question in ``objective
Bayes'' applications; see, for instance, \citet{gelman} and \citet{berger}.

Bootstrap reweighting can apply to any choice of prior (not favoring
convenience priors such as the conjugates, e.g.), but here we
will be most interested in the objective-type Bayes analyses that
dominate current practice. Jeffreys priors are featured in the
examples, more for easy presentation than necessity. The paper ends
with a brief summary in Section~\ref{sec7}. Some technical details are
deferred to the \hyperref[app]{Appendix}.

Connections between nonparametric bootstrapping and Bayesian inference
emerged early, with the ``Bayesian bootstrap,'' \citet{rubin} and \citet
{1982}. Bootstrap reweighting is deployed differently in \citet{smith},
with a nice example given in their Section 5. Sections 4 and 6 of \citet
{1998} develop bootstrap reweighting along the lines used in this paper.

\begin{table}[b]
\tabcolsep=0pt
\caption{Scores of 22 students on two tests, ``mechanics'' and
``vectors'' [from Mardia, Kent and Bibby (\protect\citeyear{mardia}),
a randomly chosen subset of the 88
students in their Table 1.2.1]. The~sample~correlation is
$\hthe=0.498$}\label{table1}
\begin{tabular*}{\tablewidth}{@{\extracolsep{\fill}}lrccccccccccccccccccccc@{}}
\hline
&\multicolumn{1}{c}{\textbf{1}}&\textbf{2}&\textbf{3}&\textbf{4}&\textbf{5}
&\textbf{6}&\textbf{7}&\textbf{8}&\textbf{9}&\textbf{10}&\textbf{11}&\textbf{12}
&\textbf{13}&\textbf{14}&\textbf{15}&\textbf{16}
&\textbf{17}&\textbf{18}&\textbf{19}&\textbf{20}&\textbf{21}&\textbf{22}\\
\hline
mech&7&44&49&59&34&46&\hphantom{0}0&32&49&52&44&36&42&\hphantom{0}5&22&18&41&48&31&42&46&63\\
vec&51&69&41&70&42&40&40&45&57&64&61&59&60&30&58&51&63&38&42&69&49&63\\
\hline
\end{tabular*}
\end{table}

\section{Conversion and reweighting}\label{sec2}

Our methodology is introduced here in terms of a simple one-parameter
problem. Table~\ref{table1} shows scores for $n=22$ students on two
tests, ``mechanics'' and ``vectors,'' having sample correlation
%
\begin{equation}\label{21}
\hthe=0.498.
\end{equation}
We wish to calculate some measure of posterior distribution for the
true underlying parameter value
%
\begin{equation}\label{22}
\theta_0=\mbox{correlation (mechanics score, vectors score)}.
\end{equation}

As in \citet{mardia}, we assume that the individual student scores
$y_i=(\mech_i,\vect_i)$ are a random sample from a bivariate normal
distribution having unknown mean vector $\mu$ and covariance matrix
$\nsig$,
%
\begin{equation}\label{23}
\by\dvtx y_i\stackrel{\mathrm{ind}} {\sim}\n_2(\mu,\nsig)
\qquad\mbox{for }i=1,2,\ldots,22
\end{equation}
with $\by=(y_1,y_2,\ldots,y_{22})$ representing the full data set. Let
$(\hmu,\hnsig)$ denote the usual maximum likelihood estimate (MLE).
Then a \textit{parametric bootstrap sample} $\by^*$ follows (\ref
{23}), with $(\hmu,\hnsig)$ replacing $(\mu,\nsig)$,
%
\begin{equation}\label{24}
\by^*\dvtx y_i^*\stackrel{\mathrm{ind}} {\sim}\n_2 (\hmu,
\hnsig )\qquad\mbox{for }i=1,2,\ldots,22.
\end{equation}

The sample correlation of $\by^*$ is a \textit{parametric bootstrap
replication} of $\hthe$, say,~$\hthe^*$. A total of $B=10\mbox{,}000$
parametric bootstrap samples $\by^*$ were independently generated
according to (\ref{24}), and the corresponding $\hthe^*$ values
calculated. We will denote them simply as
%
\begin{equation}\label{25}
\theta_1,\theta_2,\ldots,\theta_i,\ldots,
\theta_B
\end{equation}
with $\theta_i$ short for $\hthe_i^*$.

\begin{figure}

\includegraphics{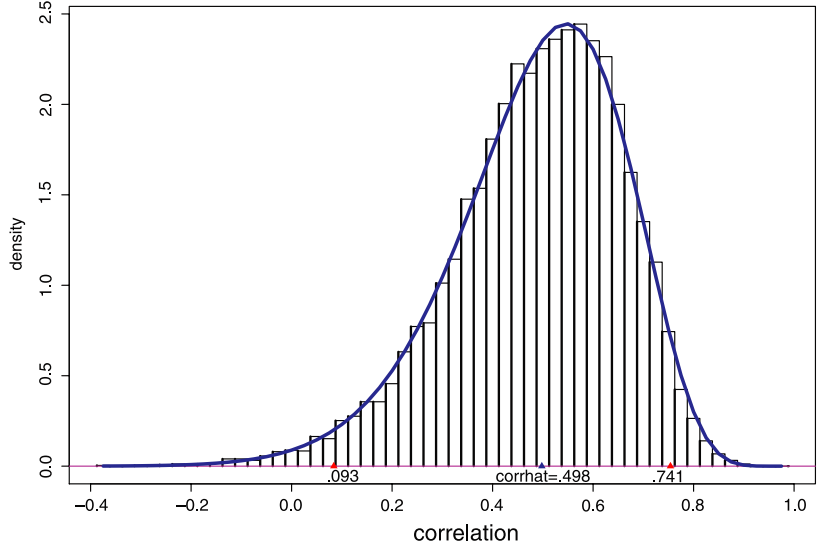}

\caption{Histogram of $B=10\mbox{,}000$ bootstrap replications for the
student score correlation coefficient (\protect\ref{24})--(\protect
\ref{25}) scaled
to integrate to 1. Solid curve is Fisher's density formula (\protect
\ref{26})
for $\theta=0.498$. Triangles indicate the exact 95\% confidence
interval $\theta\in(0.093,0.741)$.}
\label{fig1}
\end{figure}

The histogram in Figure~\ref{fig1} compares the distribution of the
10,000 $\theta_i$'s with Fisher's theoretical density function $\fthe
(\hthe)$,
%
\begin{equation}\label{26}
\fthe(\hthe)=\frac{(n-2)(1-\theta^2)^{(n-1)/2}(1-\hthe^2)^{(n-4)/2}}{\pi}\int_0^\infty
\frac{dw}{(\cosh w-\theta\hthe)^{n-1}},\hspace*{-25pt}
\end{equation}
where $\theta$ has been set equal to its MLE value 0.498. In this
sense $f_{0.498}(\cdot)$ is the ideal parametric bootstrap density we
would obtain if the number of replications $B$ approached infinity.
Chapter 32 of \citet{johnson} gives formula (\ref{26}) and other
representations of $\fthe(\hthe)$.

Figure~\ref{fig1} also indicates the exact 95\% confidence limits
%
\begin{equation}\label{27}
\theta_0\in(0.093,0.741),
\end{equation}
$2\frac12\%$ noncoverage in each tail, obtained from $\fthe(\hthe)$
by the usual construction,
%
\begin{equation}\label{28}
\int_{0.498}^1f_{0.093}(\theta)\,d
\theta=0.025
\end{equation}
and similarly at the upper endpoint.

Suppose now\setcounter{footnote}{1}\footnote{For this example we reduce the problem to finding
the posterior distribution of $\theta$ given~$\hthe$, ignoring any
information about $\theta$ in the part of $(\hmu,\hnsig)$ orthogonal
to $\hthe$. Our subsequent examples do not make such reductions.} we
have a prior density $\pi(\theta)$ for the parameter $\theta$ and
wish to
calculate the posterior density $\pi(\theta|\hthe)$. For any
subset\vadjust{\goodbreak}
$\cala$ of the parameter space $\Theta=[-1,1]$,
%
\begin{equation}\label{29}
\Pr \{\theta\in\cala|\hthe \} =\int_{\cala}\pi(\theta)\fthe(
\hthe)\,d\theta \Big/\int_\Theta \pi(\theta)\fthe(\hthe)\,d\theta
\end{equation}
according to Bayes rule.

Define the \textit{conversion factor} $R(\theta)$ to be the ratio of the
likelihood function to the bootstrap density,
%
\begin{equation}\label{210}
R(\theta)=\fthe(\hthe)/\fhthe(\theta).
\end{equation}
Here $\hthe$ is fixed at its observed value 0.498 while $\theta$
represents any point in~$\Theta$. We can rewrite (\ref{29}) as
%
\begin{equation}\label{211}
\Pr \{\theta\in\cala|\hthe \} =\frac{\int_{\cala}\pi(\theta)R(\theta)\fhthe(\theta)\,d\theta
}{\int_\Theta\pi(\theta)R(\theta)\fhthe(\theta)\,d\theta}.
\end{equation}
More generally, if $t(\theta)$ is any function $\theta$, its posterior
expectation is
%
\begin{equation}\label{212}
E \bigl\{t(\theta)|\hthe \bigr\} =\frac{\int_\Theta t(\theta)\pi(\theta)R(\theta)\fhthe(\theta)\,d\theta
}{\int_\Theta\pi(\theta)R(\theta)\fhthe(\theta)\,d\theta}.
\end{equation}

The integrals in (\ref{211}) and (\ref{212}) are now being taken with
respect to the parametric bootstrap density $\fhthe(\cdot)$. Since
$\theta_1,\theta_2,\ldots,\theta_B$ (\ref{25}) is a random sample
from~$\fhthe(\cdot)$, the integrals can be estimated by sample averages
in the usual way, yielding the familiar importance sampling estimate of
$E\{t(\theta)|\hthe\}$,
%
\begin{equation}\label{213}
\hate \bigl\{t(\theta)|\hthe \bigr\}=\sum_{i=1}^Bt_i
\pi_iR_i \Big/\sum_{i=1}^B
\pi_iR_i,
\end{equation}
where $t_i=t(\theta_i), \pi_i=\pi(\theta_i)$, and $R_i=R(\theta_i)$. Under mild regularity conditions, the law of large numbers
implies that $\hate\{t(\theta)|\hthe\}$ approaches $E\{t(\theta
)|\theta\}$
as $B\to\infty$. (The accuracy calculations of Section~\ref{sec6}
will show that in this case $B=10\mbox{,}000$ was larger than necessary for
most purposes.)

\begin{figure}

\includegraphics{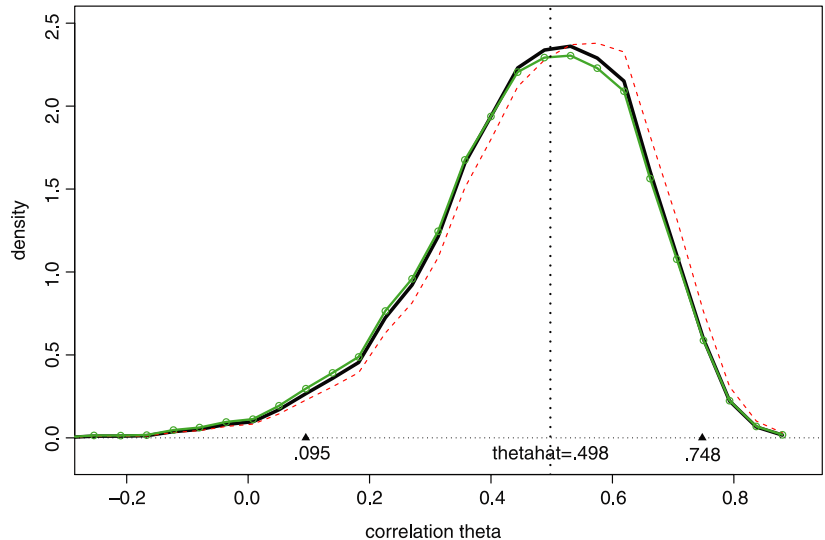}

\caption{Heavy curve is the posterior density $\pi(\theta|\hthe)$ for
the correlation (\protect\ref{22}), starting from Jeffreys prior
(\protect\ref{214}),
obtained by reweighting the $B=10\mbox{,}000$ bootstrap replications
(\protect\ref{25}); triangles show 95\% credible limits
$\theta_0\in(0.095,0.748)$. Light dashed curve is raw unweighted
bootstrap distribution. Beaded curve is BCa weighted bootstrap density
(\protect\ref{217}), nearly the same as $\pi(\hthe|\theta)$ in
this case.}
\label{fig2}
\end{figure}

The heavy curve in Figure~\ref{fig2} describes $\hpi(\theta|\hthe
)$, the estimated posterior density starting from Jeffreys prior
%
\begin{equation}\label{214}
\pi(\theta)=1/\bigl(1-\theta^2\bigr)
\end{equation}
(see Section~\ref{sec3}). The raw bootstrap distribution puts weight
$1/B$ on each of the $B$ replications $\theta_i$. By \textit
{reweighting} these points proportionately to $w_i=\pi_iR_i$, we
obtain the estimated posterior distribution of $\theta$ given $\hthe
$, with
%
\begin{equation}\label{215}
\widehat\Pr \{\theta\in\cala|\hthe \}=\sum_{\theta
_i\in\cala}w_i
\Big/\sum_{i=1}^Bw_i;
\end{equation}
$\hpi(\theta|\hthe)$ represents the density of this
distribution---essentially a smoothed histogram of the 10,000
$\theta_i$'s, weighted proportionally to $w_i$.

Integrating $\hpi(\theta|\hthe)$ yields the 95\% credible limits
($2\frac12\%$ posterior probability in each tail)
%
\begin{equation}\label{216}
\theta_0\in(0.095,0.748),
\end{equation}
close to the exact limits (\ref{27}). Prior (\ref{214}) is known to
yield accurate frequentist coverage probabilities, being a member of
the Welch--Peers family discussed in Section~\ref{sec4}.

In this case, the weights $w_i=\pi_iR_i$ can be thought of as
correcting the raw unweighted ($w_i\equiv1$) bootstrap density. Figure
\ref{fig2} shows the correction as a small shift leftward. BCa,
standing for \textit{bias-corrected and accelerated}, is another set
of corrective weights, obtained from purely frequentist considerations.
Letting $\hatg(\theta)$ denote the usual empirical cumulative
distribution function (c.d.f.) of the bootstrap replications $\theta_1,\theta_2,\ldots,\theta_B$, the BCa weight on $\theta_i$ is
%
\begin{equation}\label{217}
w_i^{\bca}=\frac{\varphi (z_{\theta i}/(1+az_{\theta
i})-z_0 )}{(1+az_{\theta i})^2\varphi(z_{\theta i}+z_0)}\qquad \bigl[z_{\theta i}=
\Phi^{-1}\hatg(\theta_i)-z_0 \bigr],
\end{equation}
where $\varphi$ and $\Phi$ are the standard normal density and c.d.f.,
while $z_0$ and $a$ are the \textit{bias-correction} and \textit
{acceleration} constants developed in \citet{1987} and \citet{1992di},
further discussed in Section~\ref{sec4} and the \hyperref[app]{Appendix}. Their
estimated values are $z_0=-0.068$ and $a=0$ for the student score
correlation problem.

The BCa density $\pi^{\bca}(\hthe|\theta)$, obtained by reweighting
as in (\ref{215}), is seen in Figure~\ref{fig2} to nicely agree with
the Jeffreys posterior density, being slightly heavier in the left
tail, with 95\% central interval $\theta_0\in(0.074,0.748)$. This
agreement is misleadingly soothing, as will be seen in the
multidimensional context of Section~\ref{sec4}.

\section{Exponential families}\label{sec3}

The Bayes/bootstrap conversion process takes on a simplified form in
exponential families. This facilitates its application to
multiparameter problems, as discussed here and in the next two sections.

The density functions for a $p$-parameter exponential family $\calf$
can be expressed as
%
\begin{equation}\label{31}
f_\beta (\hbet )=e^{\alpha'\hbet-\psi(\alpha)}f_0 (\hbet ),
\end{equation}
where the $p$-vector $\alpha$ is the canonical parameter, $\hbet$ is
the $p$-dimensional sufficient statistic vector, and where $\psi
(\alpha)$, the cumulant generating function, provides the multipliers
necessary for $f_\beta(\hbet)$ integrating to 1. Here we have indexed
the family by its expectation parameter vector $\beta$,
%
\begin{equation}\label{32}
\beta=E_\alpha \{\hbet \}
\end{equation}
for the sake of subsequent notation, but $\alpha$ and $\beta$ are
one-to-one functions and we could just as well write $f_\alpha(\hbet)$.

The \textit{deviance} between any two members of $\calf$ is
%
\begin{equation}\label{33}
D(\beta_1,\beta_2)=2E_{\beta_1} \bigl\{\log
\bigl(f_{\beta
_1} (\hbet ) /f_{\beta_2} (\hbet ) \bigr) \bigr\}
\end{equation}
[denoted equivalently $D(\alpha_1,\alpha_2)$ since deviance does not
depend on the parameterization of $\calf$]. Taking logs in (\ref{31})
shows that
%
\begin{equation}\label{34}
D(\beta_1,\beta_2)/2=(\alpha_1-
\alpha_2)'\beta_1- \bigl(\psi (
\alpha_1)-\psi(\alpha_2) \bigr).
\end{equation}
Then family (\ref{31}) can be re-expressed in ``Hoeffding's form'' as
%
\begin{equation}\label{35}
f_\beta (\hbet )=f_{\hbet} (\hbet )e^{-D(\hbet,\beta)/2}.
\end{equation}
Since $D(\hbet,\beta)$ is equal to or greater than zero, (\ref{35})
shows that $\beta=\hbet$ is the MLE, maximizing $f_\beta(\hbet)$
over all choices of $\beta$ in $\calb$, the space of possible
expectation vectors.

\textit{Parametric bootstrap replications} of $\hbet$ are independent
draws from $f_{\hbet}(\cdot)$,
%
\begin{equation}\label{36}
f_{\hbet}(\cdot)\longrightarrow\beta_1,\beta_2,
\ldots,\beta_i,\ldots,\beta_B,
\end{equation}
where $\beta_i$ is shorthand notation for $\hbet_i^*$. Starting from
a prior density $\pi(\beta)$ on $\calb$, the posterior expectation of
any function $t(\beta)$ given $\hbet$ is estimated by
%
\begin{equation}\label{37}
\hate \bigl\{t(\beta) |\hbet \bigr\}=\sum_{i=1}^Bt(
\beta_i)\pi (\beta_i)R(\beta_i) \Big/\sum
_{i=1}^B\pi(\beta_i)R(
\beta_i)
\end{equation}
as in (\ref{213}), with $R(\beta)$ the \textit{conversion factor}
%
\begin{equation}\label{38}
R(\beta)=f_\beta (\hbet ) /f_{\hbet}(\beta).
\end{equation}
\textit{Note}:
$\pi(\beta)R(\beta)$ is transformation invariant, so formula (\ref{37})
produces the same numerical result if we bootstrap $\alpha_1,\alpha_2,\ldots,\alpha_B$ instead of (\ref{36}), or for that matter
bootstrap any other sufficient vector. See Section~\ref{sec4}.

Hoeffding's form (\ref{35}) allows a convenient expression for
$R(\beta)$:
%
\begin{lemma}\label{lem31}
Conversion factor (\ref{38}) equals
%
\begin{equation}\label{39}
R(\beta)=\xi(\beta)e^{\Delta(\beta)},
\end{equation}
where
%
\begin{equation}\label{310}
\xi(\beta)=f_{\hbet} (\hbet ) /f_\beta(\beta )
\end{equation}
and
%
\begin{equation}\label{311}
\Delta(\beta)= \bigl[D (\beta,\hbet )-D (\hbet,\beta ) \bigr] /2.
\end{equation}
\end{lemma}

Letting $\halp$ be the canonical parameter vector corresponding to
$\hbet$, (\ref{34}) gives
%
\begin{equation}\label{312}
\Delta(\beta)= (\alpha-\halp )' (\beta+\hbet )-2 \bigl[\psi(
\alpha)-\psi (\halp ) \bigr],
\end{equation}
which is useful for both theoretical and numerical computations.

The derivatives of $\psi$ with respect to components of $\alpha$
yield the moments of $\hbet$,
%
\begin{equation}\label{313}
\dot\psi(\alpha)\equiv(\partial\psi/\partial\alpha_j)=\beta,\qquad\ddot
\psi (\alpha)\equiv\bigl(\partial^2\psi/\partial\alpha_j
\partial\alpha_k\bigr)=V(\alpha)=\cov_\alpha \{\hbet \}\hspace*{-32pt}
\end{equation}
and
%
\begin{equation}\label{314}
\dddot{\psi}(\alpha)\equiv\bigl(\partial^3\psi/\partial
\alpha_j\,\partial\alpha_k\,\partial\alpha_l
\bigr)=U(\alpha),
\end{equation}
$U_{jkl}(\alpha)=E_\alpha(\hbet_j-\beta_j)(\hbet_k-\beta_k)(\hbet_l-\beta_l)$. In repeated sampling situations, where $\hbet$ is
obtained from $n$ independent observations, the entries of $V(\alpha)$
and $U(\alpha)$ are typically of order $O(n^{-1})$ and $O(n^{-2})$,
respectively; see Section~5 of \citet{1987}.

The normal approximation
%
\begin{equation}\label{315}
\hbet\dotsim\n_p \bigl(\beta,V(\alpha) \bigr)
\end{equation}
yields
%
\begin{equation}\label{316}\qquad
f_\beta(\beta)\doteq(2\pi)^{-p/2}\bigl|V(\alpha)\bigr|^{-1/2}
\quad\mbox{and}\quad f_{\hbet
} (\hbet )\doteq(2\pi)^{-p/2}\bigl
\llvert V (\halp )\bigr\rrvert^{-1/2},
\end{equation}
so
%
\begin{equation}\label{317}
\xi(\beta)\doteq\bigl\llvert V(\alpha)\bigr\rrvert^{1/2} /\bigl\llvert
V (\halp )\bigr\rrvert^{1/2}.
\end{equation}
Because (\ref{316}) applies the central limit theorem where it is most
accurate, at the center, (\ref{317}) typically errs by a factor of
only $1+O(1/n)$ in repeated sampling situations; see \citet{tierney}.
In fact, for discrete families like the Poisson, where $f_\beta(\beta)$
is discontinuous, approximation (\ref{317}) yields superior
performance in applications of (\ref{39}) to (\ref{37}). In what
follows we will treat (\ref{317}) as exact rather than approximate.

Jeffreys invariant prior density, as described in \citet{kass}, takes
the form
%
\begin{equation}\label{318}
\pi^{\jeff}(\beta)=c\bigl\llvert V(\alpha)\bigr\rrvert^{-1/2}
\end{equation}
in family (\ref{31}), with $c$ an arbitrary positive constant that
does not affect estimates such as (\ref{37}). Ignoring $c$, we can use
(\ref{317}) and (\ref{318}) to rewrite the conversion factor $R(\beta)$
(\ref{39}) as
%
\begin{equation}\label{319}
R(\beta)=e^{\Delta(\beta)}/\pi^{\jeff}(\beta).
\end{equation}

Jeffreys prior is intended to be ``uninformative.'' Like other \textit
{objective} priors discussed in \citeauthor{kass}, it is designed
for\vadjust{\goodbreak}
Bayesian use in situations lacking prior experience. Its use amounts to
choosing 
%
\begin{equation}\label{320}
\pi(\beta)R(\beta)=e^{\Delta(\beta)}
\end{equation}
in which case (\ref{37}) takes on a particularly simple form:
%
\begin{lemma}\label{lem32}
If $\pi(\beta)$ is Jeffreys prior (\ref{318}), then (\ref{37}) equals
%
\begin{equation}\label{321}
\hate \bigl\{t(\beta) |\hbet \bigr\}=\sum_{i=1}^Bt(
\beta_i)e^{\Delta(\beta_i)} \Big/\sum_{i=1}^Be^{\Delta(\beta_i)}
\end{equation}
with $\Delta(\beta)$ as in (\ref{311}) and (\ref{312}).
\end{lemma}

The normal translation model $\hbet\sim\n_p(\beta,\nsig)$, with
$\nsig$ fixed, has $\Delta(\beta)=0$, so that the Bayes estimate
$\hatt$ in (\ref{321}) equals the unweighted bootstrap estimate
$\bart$,
%
\begin{equation}\label{322}
\hatt=\hate \bigl\{t(\beta) |\hbet \bigr\}=\sum_{i=1}^Bt_i\Big/B=
\bart.
\end{equation}
Usually though, $\hatt$ will not equal $\bart$, the difference
relating to the variability of $\Delta(\beta)$ in (\ref{321}).

A simple but informative result concerns the \textit{relative Bayesian
difference} (RBD) of $t(\beta)$ defined to be
%
\begin{equation}\label{323}
\rbd(t)= (\hatt-\bart ) /\barsd(t),
\end{equation}
$\barsd(t)=[\sum_1^B(t_i-\bart)^2/B]^{1/2}$:
%
\begin{lemma}\label{lem33}
Letting $r_i=\pi_iR_i$, the relative Bayesian difference of $t(\beta
)$ is
%
\begin{equation}\label{324}
\rbd(t)=\barcor(t,r)\cdot\barcv(r)
\end{equation}
and if $\pi(\beta)=\pi^{\jeff}(\beta)$,
%
\begin{equation}\label{325}
\rbd(t)\doteq\barcor(t,r)\cdot\barsd(\Delta);
\end{equation}
here $\barcor(t,r)$ is the empirical correlation between $t_i$ and
$r_i$ for the $B$ bootstrap replications, $\barcv(r)$ the empirical
coefficient of variation of the $r_i$ values, and $\barsd(\Delta)$
the empirical standard deviation of the $\Delta_i$ values.
\end{lemma}
\begin{pf}
Equation (\ref{324}) follows immediately from (\ref{37}),
%
\begin{equation}\label{326}
\rbd(t)=\frac{\sum_1^B(t_i-\bart)r_i/B}{\barsd(t)\sum_1^Br_i/B}=\barcor(t,r)\frac{\barsd(r)}{\bar{r}}.
\end{equation}
If $\pi(\beta)$ is the Jeffreys prior (\ref{318}), then $r(\beta
)=\exp
(\Delta(\beta))$ (\ref{319}) and the usual delta-method argument gives
$\barcv(r)\doteq\barsd(\Delta)$.
\end{pf}

The student score example of Figure~\ref{fig2} (which is not in
exponential family form) has, directly from definition (\ref{323}),
%
\begin{equation}\label{327}
\rbd(t)=\frac{0.473-0.490}{0.169}=-0.101,
\end{equation}
which is also obtained from (\ref{324}) with $\barcor(t,r)=-0.945$ and
$\barcv(r)=0.108$. Notice that the $\barcv(r)$ factor in (\ref {324}),
and likewise $\barsd(\Delta)$ in (\ref{325}), apply to \textit{any}
function $t(\beta)$, only the $\barcor(t,r)$ factor being particular.
The multiparameter examples of Sections~\ref{sec3} and~\ref{sec4} have
larger $\barcv(r)$ but smaller $\barcor(t,r)$, again yielding rather
small values of $\rbd(t)$. All of the Jeffreys prior examples in this
paper show substantial agreement between the Bayes and unweighted
bootstrap results.

Asymptotically, the deviance difference $\Delta(\beta)$ depends on the
skewness of the exponential family. A normal translation family has
zero skewness, with $\Delta(\beta)=0$ and $R(\beta)=1$, so the
unweighted parametric bootstrap distribution is the same as the
flat-prior Bayes posterior distribution. In a repeated sampling
situation, skewness goes to zero as $n^{-1/2}$, making the Bayes and
bootstrap distributions converge at this rate. We can provide a simple
statement in one-parameter families:
%
\begin{theorem}\label{thm34}
In a one-parameter exponential family, $\Delta(\beta)$ has the Taylor
series approximation
%
\begin{equation}\label{328}
\Delta(\beta)\doteq\tfrac16\hgam Z^3\qquad \bigl[Z=\hatv^{-1/2} (
\beta-\hbet ) \bigr],
\end{equation}
where $\hatv$ and $\hgam$ are the variance and skewness of $\beta
\sim f_{\hbet}(\cdot)$. In large-sample situations, $Z\dotsim\n
(0,1)$ and $\hgam$ is $O(n^{-1/2})$, making $\Delta(\beta)$ of order
$O_p(n^{-1/2})$.
\end{theorem}

(The proof appears in the \hyperref[app]{Appendix}, along with the theorem's
multiparameter version.)

As a simple example, suppose
%
\begin{equation}\label{329}
\hbet\sim\beta\mathrm{Gamma}_n/n\qquad \bigl[\beta\in(0,\infty) \bigr],
\end{equation}
so $\hbet$ is a scaled version of a standard Gamma variate having $n$
degrees of freedom. In this case,
%
\begin{equation}\label{330}
\Delta(\beta)\doteq\frac1{3\sqrt{n}}Z^3\qquad\mbox{with }Z=\sqrt {n}
\biggl(\frac{\beta}{\hbet}-1 \biggr),
\end{equation}
making $\Delta(\beta)$ an increasing cubic function of $\beta$. The
cubic nature of (\ref{328}) and (\ref{330}) makes reweighting of the
parametric bootstrap replications $\beta_i$ by $\exp(\Delta_i)$ more
extreme in the tails of the distribution than near $\hbet$.

Stating things in terms of conditional expectations $\hate\{t(\beta)
|\hbet\}$ as in (\ref{37}) is convenient, but partially obscures the
basic idea: that the distribution putting\vspace*{1pt} weight proportional to
$w_i=\pi_iR_i$ on $\beta_i$ approximates the posterior distribution
$\pi(\beta|\hbet)$.\vadjust{\goodbreak}

As an example of more general Bayesian calculations, consider the
``posterior predictive distribution,''
%
\begin{equation}\label{331}
g(\by)=\int\pi (\beta |\hbet )g_\beta(\by)\,d\beta,
\end{equation}
where\vspace*{1pt} $\by$ is the original data set yielding $\hbet$; by sufficiency
as in (\ref{23}), it has density functions $g_\beta(\by)=f_\beta
(\hbet)h(\by|\hbet)$. For each $\beta_i$ we sample $\by_i^{**}$
from $g_{\beta_i}(\cdot)$. Then the discrete distribution putting
weight proportional to $w_i$ on $\by_i^{**}$, for $i=1,2,\ldots,B$,
approximates $g(\by)$. See \citet{gelman}.

\section{The multivariate normal family}\label{sec4}

This section and the next illustrate Bayes/bootstrap relationships in
two important exponential families: the multivariate normal and
generalized linear models. A multivariate normal sample $\by$
comprises $n$ independent $d$-dimensional normal vector observations
%
\begin{equation}\label{41}
\by\dvtx y_i\stackrel{\mathrm{ind}} {\sim}\n_d(\mu,\nsig),
\qquad i=1,2,\ldots,n.
\end{equation}
This involves $p=d\cdot(d+3)/2$ unknown parameters, $d$ for the mean
vector $\mu$ and $d\cdot(d+1)/2$ for the covariance matrix $\nsig$.
We will use $\gamma$ to denote the vector of all $p$ parameters;
$\gamma$ is \textit{not} the expectation vector $\beta$ (\ref{32}),
but rather a one-to-one quadratic function $\gamma=m(\beta)$
described in formula (3.5) of \citet{1992di}.

The results of Section~\ref{sec3} continue to hold under smooth
one-to-one transformations $\gamma=m(\beta)$. Let $\tilf_\gamma
(\hgam)$ denote the density of the MLE $\hgam=m(\hbet)$, and
likewise $\tilr(\gamma)=\tilf_\gamma(\hgam)/\tilf_{\hgam}(\gamma)$
for the conversion factor, $\tild(\gamma_1,\gamma_2)$ for the
deviance, $\tildel(\gamma)=[\tild(\gamma,\hgam)-\tild(\hgam,\gamma
)]/2$ for the deviance difference, and $\tilpi^{\jeff}(\gamma)$ for
Jeffreys prior. Then Lemma~\ref{lem31} continues to apply in the
transformed coordinates:
%
\begin{equation}\label{42}
\tilr(\gamma)=\tilxi(\gamma)e^{\tildel(\gamma)}\qquad \bigl[\tilxi(\gamma) =
\tilf_{\hgam} (\hgam ) /\tilf_\gamma(\gamma ) \bigr].
\end{equation}
(See the \hyperref[app]{Appendix}.)

A parametric bootstrap sample
%
\begin{equation}\label{43}
\tilf_\gamma(\cdot)\longrightarrow\gamma_1,
\gamma_2,\ldots,\gamma_B
\end{equation}
approximates the conditional expectation of a function $\tilt(\gamma)$,
starting from prior $\tilpi(\gamma)$, by
%
\begin{equation}\label{44}
\hate \bigl\{\tilt(\gamma) |\hgam \bigr\}=\sum_{i=1}^B
\tilt_i\tilpi_i\tilr_i \Big/\sum
_{i=1}^B\tilpi_i\tilr_i
\end{equation}
as in (\ref{214}), and if $\tilpi(\gamma)$ is Jeffreys prior,
%
\begin{equation}\label{45}
\hate \bigl\{\tilt(\gamma) |\hgam \bigr\}=\sum_{i=1}^B
\tilt_ie^{\tildel_i} \Big/\sum_{i=1}^Be^{\tildel_i}
\end{equation}
as in (\ref{321}). This can be particularly handy since $\Delta$ is
tranformation invariant and can be evaluated in any convenient set of
coordinates, while $\tilpi^{\jeff}(\gamma)$ need not be calculated
at all.

The following theorem provides $\tilxi(\gamma)$ and $\tilr(\gamma)$
for a
multivariate normal sample (\ref{41}), working with $\gamma$ the
$p=d\cdot(d+3)/2$ coordinates consisting of $\mu$ and the elements of
$\nsig$ on or above its main diagonal:
%
\begin{theorem}\label{thm41}
In $(\mu,\nsig)$ coordinates,
%
\begin{equation}\label{46}
\tilxi(\mu,\nsig)= \bigl(|\nsig| /|\hnsig| \bigr)^{
({d+2})/2}
\end{equation}
and
%
\begin{eqnarray}\label{47}
\tildel(\mu,\nsig)&=&n \biggl\{ (\mu-\hmu )'\frac{\hnsig^{-1}-\nsig^{-1}}2 (\mu-
\hmu )\nonumber\\[-8pt]\\[-8pt]
&&\hspace*{10.5pt}{}+\frac{\tr(\nsig\nsig^{n-1}-\hnsig\nsig^{-1})}2+\log\frac{|\hnsig|}{|\nsig|}
\biggr\}.\nonumber
\end{eqnarray}
\end{theorem}

(Proof in the \hyperref[app]{Appendix}.)

Here $1/\tilxi(\mu,\nsig)$ turns out to be exactly proportional to
$|\tilde{V}(\gamma)|^{-1/2}$, and either expression gives $\tilpi^{\jeff}
(\mu,\nsig)$. Expression (\ref{47}) equals the deviance difference
(\ref
{311}), no matter what the choice of coordinates.

Theorem~\ref{thm41} makes\vspace*{1pt} it easy to carry out parametric
bootstrapping: having calculated the usual MLE estimates $(\hmu,\hnsig
)$, each bootstrap data set $\by^*$ is generated as in (\ref{41}),
%
\begin{equation}\label{48}
\by^*\dvtx y_i^*\sim\n_d (\hmu,\hnsig ),\qquad i=1,2,
\ldots,n,
\end{equation}
from which we calculate the bootstrap MLE estimate $(\hmu^*, \hnsig^*)$, denoted simply $(\mu,\nsig)$ as before. To each of $B$ such replicates
%
\begin{equation}\label{49}
(\mu,\nsig)_1,(\mu,\nsig)_2,\ldots,(\mu,
\nsig)_i,\ldots,(\mu,\nsig)_B
\end{equation}
is attached the weight
%
\begin{equation}\label{410}
w_i=\tilpi_i\tilxi_ie^{\tildel_i}
\end{equation}
using Theorem~\ref{thm41} (or more exactly $w_i/\sum_1^Bw_j$); this
distribution, supported on the $B$ points (\ref{49}), estimates the
posterior distribution of $(\mu,\nsig)$ given $(\hmu,\hnsig)$.
Expectations are then obtained as in (\ref{44}), and similarly for
more general posterior parameters such as percentiles and credible limits.

\begin{figure}

\includegraphics{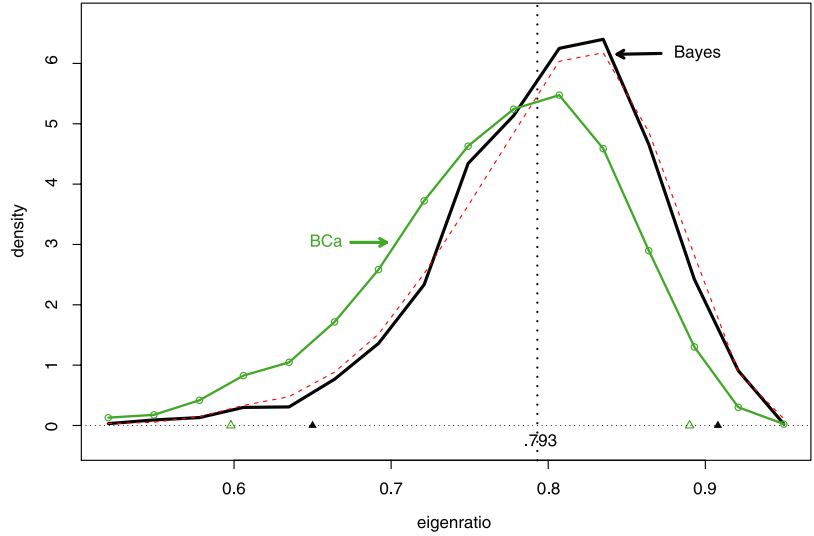}

\caption{\textup{Heavy curve} is Bayes posterior density for the
eigenratio (\protect\ref{411}), starting from Jeffreys prior for a bivariate
normal model; solid triangles show 95\% credible limits
$(0.650,0.908)$. \textup{Beaded curve} is BCa confidence density based
on weights (\protect\ref{217}) with $z_0=-0.222, a=0$; BCa 95\% interval
$(0.598,0.890)$, open triangles, is shifted far leftward. \textup{Light
dashed curve} is unweighted bootstrap density.} \label{fig3}
\end{figure}

Figure~\ref{fig3} applies this methodology to the student score data
of Table~\ref{table1}, assuming the bivariate normal model (\ref
{23}). We take the parameter of interest $\theta$ to be the \textit
{eigenratio}
%
\begin{equation}\label{411}
\theta=t(\mu,\nsig)=\lambda_1/(\lambda_1+
\lambda_2),
\end{equation}
where $\lambda_1$ and $\lambda_2$ are the ordered eigenvalues of
$\nsig$; $\theta$ has MLE $\hthe=t(\hmu,\hnsig)=0.793$.\vadjust{\goodbreak}

$B=10\mbox{,}000$ bootstrap replications were generated as in (\ref{49}), and
$t_i=t((\mu,\allowbreak\nsig)_i)$ calculated for each. Total computation time was
about 30 seconds. The heavy curve shows the estimated posterior density
of $\theta$ given $(\hmu,\hnsig)$, starting from Jeffreys prior. The
95\% credible region, $2\frac12\%$ probability excluded in each tail, was
%
\begin{equation}\label{412}
\mbox{\textit{Bayes}:}\quad\theta\in(0.650,0.908).
\end{equation}
That is,
%
\begin{equation}\label{413}
\sum_{t_i\leq0.650}e^{\tildel_i} \Big/\sum
_1^Be^{\tildel_i}=0.025
\end{equation}
and similarly for the upper endpoint.

In this case the BCa 95\% confidence limits are shifted sharply
leftward compared to (\ref{412}),
%
\begin{equation}\label{414}
\mbox{\textit{BCa}:}\quad\theta\in(0.598,0.890).
\end{equation}
The beaded curve in Figure~\ref{fig3} shows the full BCa confidence
density, that is, the estimated density based on the BCa weights (\ref
{217}). For the eigenratio, $z_0=-0.222$ and $a=0$ are the bias
correction and acceleration constants. See the \hyperref[app]{Appendix} for a brief
discussion of the $z_0$ and $a$ calculations.

\begin{figure}

\includegraphics{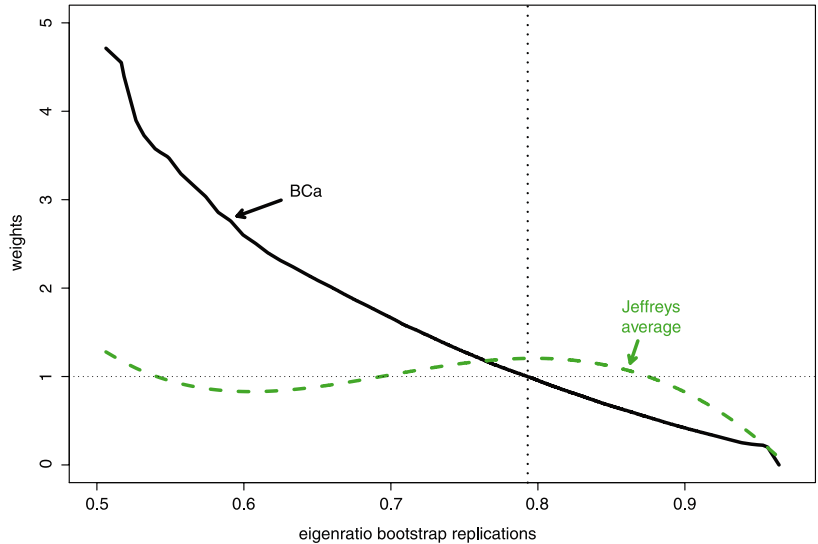}

\caption{\textup{Solid curve}: BCa weights (\protect\ref{217}), with
$(z_0,a)=(-0.222,0)$, plotted versus bootstrap eigenratio replications
$\theta_i$. \textup{Dashed curve}: regression of Jeffreys prior Bayes
weights $\exp(\tildel_i)$ on $\theta_\lambda$.}
\label{fig4}
\end{figure}

Figure~\ref{fig4} helps explain the difference between the Bayes and
BCa results. The heavy curve shows the BCa weights (\ref{217})
increasing sharply to the left as a function of $\theta_i=t((\mu,\nsig)_i)$, the bootstrap eigenratio values. In other words, smaller values
of $\theta_i$ are weighted more heavily, pulling the weighted
percentile points of the BCa distribution downward. On the other hand,
the Bayes weights $\tilpi^{\jeff}_i\tilr_i=\exp(\tildel_i)$ (represented
in Figure~\ref{fig4} by their regression on $\theta_i$) are nearly
flat, so that the Bayes posterior density is almost the same as the
unweighted bootstrap density shown in Figure~\ref{fig3}.

The BCa limits are known to yield highly accurate coverage
probabilities; see \citet{1996}. Moreover, in the eigenratio case, the
MLE $\hthe$ is strongly biased upward, suggesting a downward shift for
the confidence limits. This brings up a familiar complaint against
Jeffreys priors, extensively discussed in \citet{ghosh}: that in
multiparameter settings they can give inaccurate inferences for
individual parameters of interest.

This is likely to be the case for any general-purpose recipe for
choosing objective prior distributions in several dimensions. For
instance, repeating the eigenratio analysis with a standard inverse
Wishart prior on $\nsig$ (covariance matrix $I$, degrees of freedom 2)
and a flat prior on $\mu$ gave essentially the same results as in
Figure~\ref{fig3}. Specific parameters of interest require
specifically tailored priors, as with the Bernardo--Berger
\textit{reference priors}, again nicely reviewed by \citet{ghosh}.

In fact, the BCa weights can be thought of as providing such tailoring:
define the \textit{BCa prior} (relative to the unweighted bootstrap
distribution) to be
%
\begin{equation}\label{415}
\pi_i^{\bca}=w_i^{\bca}
/R_i
\end{equation}
with $w_i^{\bca}$ as in (\ref{217}). This makes the posterior weights
$\pi_i^{\bca}R_i$ appearing in expressions like (\ref{37}) equal the
BCa weights $w_i^{\bca}$, and makes posterior credible limits based on
the $\pi^{\bca}$ prior equal BCa limits. Formula (\ref{415}) can be
thought of as an automatic device for constructing \citeauthor{welch}'
(\citeyear{welch})
``probability matching priors;'' see \citet{tibs}.

Importance sampling methods such as (\ref{45}) can suffer from
excessive variability due to occasional large values of the weights.
The ``internal accuracy'' formula (\ref{62}) will provide a warning of
numerical problems. A variety of helpful counter-tactics are available,
beginning with a simple truncation of the largest weight.

Variations in the parametric bootstrap sampling scheme can be employed.
Instead of (\ref{36}), for instance, we might obtain
$\beta_1,\beta_2,\ldots,\beta_B$ from
%
\begin{equation}\label{416}
\beta_i\stackrel{\mathrm{ind}} {\sim}\n_p \bigl(
\hmu_\beta,h (\hnsig_\beta ) \bigr),
\end{equation}
where $\hmu_\beta$ and $\hnsig_\beta$ are the observed mean and
covariance of $\beta$'s from a preliminary $f_{\hbet}(\cdot)$ sample.
Here $h(\hnsig_\beta)$ indicates an expansion of $\hnsig_\beta$
designed to broaden the range of the bootstrap distribution, hence
reducing the importance sampling weights. If a regression analysis of
the preliminary sample showed the weights increasing in direction $v$
in the $\beta$ space, for example, then $h(\hnsig_\beta)$ might
expand $\hnsig_\beta$ in the $v$ direction. Devices such as this
become more necessary in higher-dimensional situations, where extreme
variability of the conversion factor $R(\beta_i)$ may destabilize our
importance sampling computations.

Replacing (\ref{36}) with (\ref{416}) changes the conversion factor
$R(\beta)$ (\ref{38}), but in an easily computable way. In fact,
replacing (\ref{36}) with $\beta_i\sim\n_p(\hmu_\beta,\hnsig_\beta)$ makes the calculation of $R(\beta)$ easier in situations where
there is no simple formula for the bootstrap density $f_{\hbet}(\beta)$.

\section{Generalized linear models}\label{sec5}

The Bayes/bootstrap conversion theory of Section~\ref{sec3} applies
directly to generalized linear models (GLM). A GLM begins with a
one-parameter exponential family
%
\begin{equation}\label{51}
g_\eta(y)=e^{\eta y-\phi(\eta)}g_0(y),
\end{equation}
where $\eta=\alpha, y=\hbet$, and $\phi(\eta)=\psi(\alpha)$ in
notation (\ref{31}). An $n\times p$ structure matrix $X$ and a
$p$-dimensional parameter vector $\alpha$ then yield an $n$-vector
$\bmet=X\alpha$, with each entry $\eta_j$ governing an independent
observation $y_j$,
%
\begin{equation}\label{52}
y_j\stackrel{\mathrm{ind}} {\sim}g_{\eta_j}(\cdot)\qquad
\mbox{for }j=1,2,\ldots,n.
\end{equation}

All of this results in a $p$-parameter exponential family (\ref{31}),
with $\alpha$ the canonical parameter vector. Letting $\bmu$ be the
expectation vector of $\by=(y_1,\ldots,y_n)'$,
%
\begin{equation}\label{53}
\bmu=E_\alpha\{\by\},
\end{equation}
the other entries of (\ref{31}) are
%
\begin{equation}\label{54}
\hbet=X'\by,\qquad \beta=X'\bmu\quad\mbox{and}\quad\psi(
\alpha )=\sum_{i=1}^n\phi(
\bx_j\alpha),
\end{equation}
where $\bx_j$ is the $j$th row of $X$. The deviance difference $\Delta
(\beta)$ (\ref{311}) has a simple form,
%
\begin{eqnarray}
\label{55} \Delta(\beta)&=& (\alpha-\halp )' (\beta+\hbet )-2\sum
_{j=1}^n \bigl[\phi(\bx_j
\alpha)-\phi (\bx_j\halp ) \bigr]
\nonumber\\[-8pt]\\[-8pt]
&=& (\bmet-\hbmet )' (\bmu+\hbmu )-2\sum
_{j=1}^n \bigl[\phi(\bmet_j)-\phi (
\hbmet_j ) \bigr]
\nonumber
\end{eqnarray}
[$\halp$ the MLE of $\alpha$, $\hbmet=X\halp$, and $\hbmu$ the
expectation vector (\ref{53}) corresponding to $\alpha=\halp$]
according to (\ref{312}).

As an extended example we now consider a microarray experiment
discussed in \citet{2010}, Section 2.1: 102 men, 50 healthy controls
and 52 prostate cancer patients, having each had the activity of
$N=6033$ genes measured [\citet{singh}]. A two-sample test comparing
patients with controls has been performed for each gene, yielding a
\textit{$z$-value} $z_k$, that is, a test statistic having a standard
normal distribution under $H_{0k}$, the null hypothesis of no
patient/control difference for gene~$k$,
%
\begin{equation}\label{56}
H_{0k}\dvtx z_k\sim\n(0,1).
\end{equation}
The experimenters, of course, are interested in identifying nonnull genes.

\begin{figure}

\includegraphics{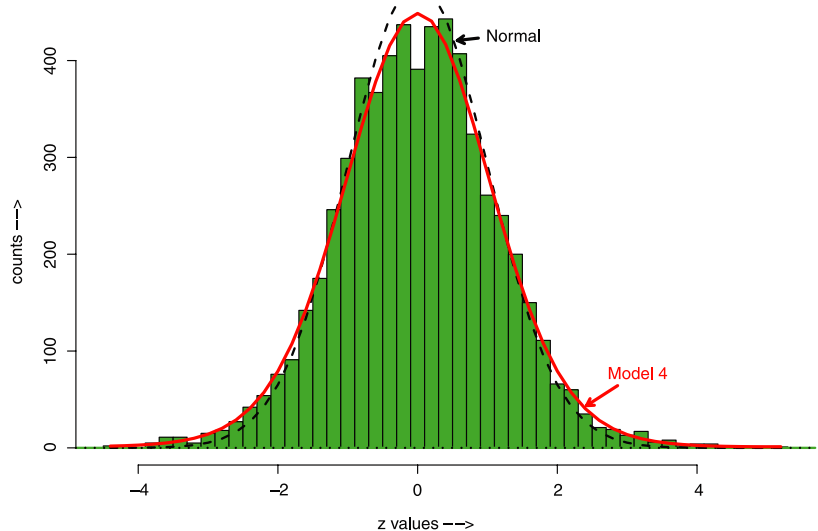}

\caption{Histogram of the $N=6033$ $z$-values from the prostate cancer
study, Singh et~al. (\protect\citeyear{singh}). Standard normal curve (dashed) is too high at
center and too low in the tails. ``Model 4,'' solid curve, is the fit
from a fourth-degree polynomial Poisson regression.}
\label{fig5}
\end{figure}

Figure~\ref{fig5} shows a histogram of the $N$ $z$-values. The standard
normal curve is too high in the center and too low in the tails,
suggesting that at least some of the genes are nonnull. The
better-fitting curve ``model 4'' is a fit from the Poisson regression
family discussed next.

There are $J=49$ bins for the histogram, each of width 0.2, with
centers $x_j$ ranging from $-4.4$ to 5.2. Let $y_j$ be the number of
$z_k$ values in the $j$th bin,
%
\begin{equation}\label{57}
y_j=\#\{z_k\in\operatorname{bin}j\},\qquad j=1,2,\ldots,
J=49.
\end{equation}
We will assume that the $y_j$'s are independent Poisson observations,
each having its own expectation $\mu_j$,
%
\begin{equation}\label{58}
y_j\stackrel{\mathrm{ind}} {\sim}\poi(\mu_j),\qquad
j=1,2,\ldots,J,
\end{equation}
and then fit curves to the histogram using Poisson regression. Why this
might be appropriate is discussed at length in Efron (\citeyear{2008,2010}), but
here we will just take it as a helpful example of the Bayes/bootstrap
GLM modeling theory.

We consider Poisson regression models where the canonical parameters
$\eta_j=\log(\mu_j)$ are $m$th-degree polynomial functions of the
bin centers $x_j$, evaluated by \texttt{glm(y$\sim
$poly(x,m),Poisson)} in the language R. This is a GLM with the Poisson
family, $\eta_j=\log\mu_j$, where $X$ is a $J\times(m+1)$ matrix
having rows $\bx_j=(1,x_j,x_j^2,\ldots,x_j^m)$ for $j=1,2,\ldots,J$.
For the Poisson distribution, $\phi(\eta)=\mu$ in (\ref{51}). The
deviance difference function (\ref{55}) becomes
%
\begin{equation}\label{59}
\Delta(\beta)= (\bmet-\hbmet )' (\bmu+\hbmu )-2\cdot
\bone' (\bmu-\hbmu )
\end{equation}
with $\bone$ a vector of $J$ ones.

\begin{table}
\caption{Deviance from Poisson polynomial regression models for counts
(\protect\ref{57}), prostate data; AIC~criterion (\protect\ref
{510}) is minimized for
the quartic model M4. \textup{Boot} \% shows the proportion of each
model selected in $B=4000$ bootstrap replications of the AIC criterion,
bootstrapping from M8. \textup{Bayes} \% are weighted Bayes posterior
proportions, assuming~Jeffreys prior. The \textup{St Error} column is
obtained from the bootstrap-after-bootstrap calculations of Section~\protect\ref{sec6}}
\label{table2}
\begin{tabular*}{\tablewidth}{@{\extracolsep{\fill}}ld{3.1}rd{3.0}d{3.0}c@{}}
\hline
\textbf{Model}&\multicolumn{1}{c}{\textbf{Deviance}}&\multicolumn{1}{c}{\textbf{AIC}}
&\multicolumn{1}{c}{\hspace*{-2pt}\textbf{Boot \%}}&\multicolumn{1}{c}{\hspace*{-4pt}\textbf{Bayes \%}}
&\multicolumn{1}{c@{}}{\textbf{(St Error)}}\\\hline
M2&138.6&144.6&0\%&0\%&\hphantom{0}(0\%)\\
M3&137.1&145.1&0\%&0\%&\hphantom{0}(0\%)\\
M4&65.3&\textbf{75.3}&32\%&36\%&(20\%)\\
M5&64.3&76.3&10\%&12\%&(14\%)\\
M6&63.8&77.8&5\%&5\%&\hphantom{0}(8\%)\\
M7&63.8&79.8&1\%&2\%&\hphantom{0}(6\%)\\
M8&59.6&77.6&51\%&45\%&(27\%)\\
\hline
\end{tabular*}
\end{table}

Let ``M$m$'' indicate the Poisson polynomial regression model of degree~$m$.
M2, with $\log(\mu_j)$ quadratic in $x_j$, amounts to a normal
location-scale model for the marginal density of the $z_k$'s.
Higher-order models are more flexible. M4, the quartic model, provided
the heavy fitted curve in Figure~\ref{fig5}. Table~\ref{table2} shows
the Poisson deviance for the fitted models M2 through M8. A dramatic
decrease occurs between M3 and M4, but only slow change occurs after
that. The AIC criterion for model $m$,
%
\begin{equation}\label{510}
\operatorname{AIC}(m)=\mathrm{Deviance}+2\cdot(m+1)
\end{equation}
is minimized at M4, though none of the subsequent models do much worse.
The fit from M4 provided the ``model 4'' curve in Figure~\ref{fig5}.

Parametric bootstrap samples $\by^*$ were generated from M4, as in
(\ref{58}),
%
\begin{equation}\label{511}
\by_j^*\stackrel{\mathrm{ind}} {\sim}\poi (\hmu_j )
\qquad\mbox{for }j=1,2,\ldots,J
\end{equation}
with $\hmu_j$ the MLE values from M4. $B=4000$ such samples were
generated, and for each one the MLE $\halp^*$, and also $\hbet^*$
(\ref{54}), were obtained from the R call \texttt{glm(y$^*\sim
$poly(x,4),Poisson)}. Using the simplified notation $\alpha=\halp^*$
gives bootstrap vectors $\bmet=X\alpha, \bmu=\exp(\bmet)=(\exp
(\eta_j)), \beta=X'\mu$, where $X$ is the $49\times5$ matrix
\texttt{poly(x,4)}, and finally $\Delta(\beta)$ as in (\ref{59}).
[Notice that $\beta$ represents $\hbet^*$ here, not the ``true
value'' $\beta$ of (\ref{54}).]

The reweighted bootstrap distribution, with weights proportional to
%
\begin{equation}\label{512}
w_i=e^{\Delta_i}\mbox{ on }\beta_i\qquad\mbox{for
}i=1,2,\ldots,B=4000,
\end{equation}
estimates the posterior distribution of $\beta$ given $\beta_i$,
starting from Jeffreys prior. The posterior expectation of any
parameter $\theta=t(\beta)$ is estimated by $\sum w_it_i/\sum w_i$ as
in (\ref{321}).

We will focus attention on a false discovery rate (Fdr) parameter
$\theta$,
%
\begin{equation}\label{513}
\theta(z)=\fdr(z)= \bigl[1-\Phi(z) \bigr] / \bigl[1-F(z) \bigr],
\end{equation}
where $\Phi$ is the standard normal c.d.f. and $F(z)$ is the c.d.f. of the
Poisson regression model: in terms of the discretized situation (\ref{58}),
%
\begin{equation}\label{514}
F(z)=\sum_{x_j\leq z}\mu_j \Big/\sum
_1^J\mu_j
\end{equation}
(with a ``half count'' correction at $z=x_j$). $\fdr(z)$ estimates the
probability that a gene having its $z_k$ exceeding the fixed value $z$
is nonnull, as discussed, for example, in \citet{2008}.

\begin{figure}

\includegraphics{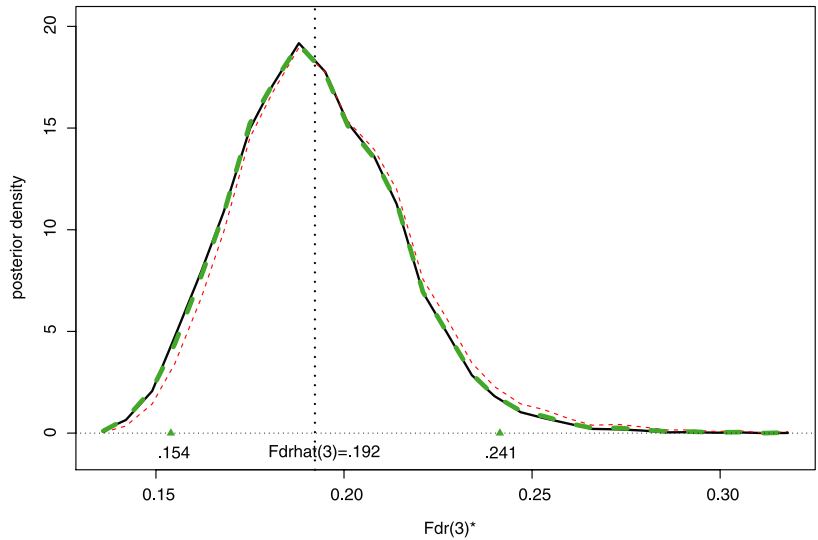}

\caption{Posterior densities for $\theta=\fdr(3)$ (\protect\ref
{513}), prostate
data, based on $B=4000$ parametric bootstrap replications (\protect
\ref{511})
from the fourth-degree Poisson regression model M4. \textup{Solid
curve} Jeffreys Bayes posterior density, using (\protect\ref{512});
\textup{heavy dashed curve} BCa confidence density (\protect\ref{217}).
Both give 95\% interval $\theta\in(-0.154,-0.241)$. \textup{Light
dashed curve} is unweighted bootstrap density. Total computation time
was about 30 seconds.} \label{fig6}
\end{figure}

Figure~\ref{fig6} concerns the choice $z=3$. Using quartic model M4 to
estimate the $\mu_j$'s in (\ref{514}) yields point estimate
%
\begin{equation}\label{515}
\hthe=\hfdr(3)=0.192.
\end{equation}
Fdr values near 0.2 are in the ``interesting'' range where the gene
might be reported as nonnull, making it important to know the accuracy
of (\ref{515}).

The $B=4000$ bootstrap samples for M4 (\ref{511}) yielded bootstrap
replications $\theta_1,\theta_2,\ldots,\theta_B$. Their standard
deviation is a bootstrap estimate of standard error for $\hthe, \hse
=0.024$, so a typical empirical Bayes analysis might report $\hfdr
(3)=0.0192\pm0.024$. A Jeffreys Bayes analysis gives the full
posterior density of $\theta$ shown by the solid curve in Figure \ref
{fig6}, with 95\% credible interval
%
\begin{equation}\label{516}
\mbox{M4:}\quad\theta\in(0.154,0.241).
\end{equation}
In this case the BCa density (\ref{217}) [$(z_0,a)=(-0.047,-0.026)$]
is nearly the same as the Bayes estimate, both of them lying just
slightly to the left of the unweighted bootstrap density.

\begin{figure}

\includegraphics{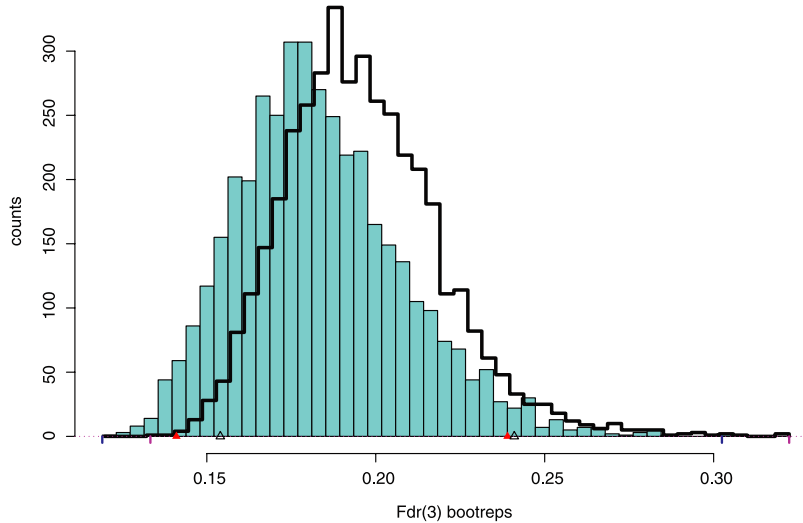}

\caption{$B=4000$ parametric bootstrap replications of $\fdr(3)$ from
M8 (solid histogram) compared with those from M4 (line histogram).
Closed triangles indicate 95\% M8 credible limits $(0.141,0.239)$; open
triangles M4 limits $(0.154,0.241)$.}
\label{fig7}
\end{figure}

The choice of philosophy, Jeffreys Bayes or BCa frequentist, does not
make much difference here, but the choice of model does. Repeating the
analysis using M8 instead of M4 to generate the bootstrap samples
(\ref{511}) sharply decreased the estimate. Figure~\ref{fig7}
compares the bootstrap histograms; the 95\% credible interval for $\fdr
(3)$ is now
%
\begin{equation}\label{517}
\mbox{M8:}\quad\theta\in(0.141,0.239).
\end{equation}

AIC calculations were carried out for each of the 4000 M8 bootstrap
samples. Of these, 32\% selected M4 as the minimizer, compared with
51\% for M8, as shown in the \textit{Boot \%} column of Table \ref
{table2}. Weighting each sample proportionally to $\exp(\Delta_i)$
(\ref{512}) narrowed the difference to 36\% versus 45\%, but still
with a strong tendency toward M8.

It might be feared that M8 is simply justifying itself. However,
standard nonparametric bootstrapping (resampling the $N$ $z_k$ values)
gave slightly \textit{more} extreme Boot percentages,
%
\begin{equation}\label{518}
\mbox{30\%(M4), 9\%(M5), 4\%(M6),
2\%(M7), 54\%(M8)}.
\end{equation}
The fact is that data-based model selection is quite unstable here, as
the accuracy calculations of Section~\ref{sec6} will verify.

\section{Accuracy}\label{sec6}

Two aspects of our methodology's Bayesian estimation accuracy are
considered in this section: \textit{internal accuracy}, the bootstrap
sampling error in estimates such as (\ref{37}) (i.e., how many
bootstrap replications $B$ need we take?),\vadjust{\goodbreak} and \textit{external
accuracy}, statistical sampling error, for instance, how much would the
results in Figure~\ref{fig3} change for a new sample of 22 students?
The i.i.d. (independent and identically distributed) nature of
bootstrap sampling makes both questions easy to answer.

Internal accuracy is particularly straightforward. The estimate (\ref
{37}) for $\hate\{t(\beta)|\allowbreak \hbet\}$ can be expressed in terms of
$s_i=t_i\pi_iR_i$ and $r_i=\pi_iR_i$ as
%
\begin{equation}\label{61}
\hate=\bars/\barr\qquad \Biggl(\bars=\sum_1^Bs_i/B,
\barr=\sum_1^Br_i/B
\Biggr).
\end{equation}
Let $\barcov$ be the $2\times2$ empirical covariance matrix of the
$B$ vectors $(s_i,r_i)$. Then standard delta-method calculations yield
a familiar approximation for the bootstrap coefficient of variation of
$\hate$,
%
\begin{equation}\label{62}
\hcv^2=\frac1{B} \biggl(\frac{\barc_{ss}}{\bars^2}-2\frac{\barc_{sr}}{\bars\barr}+
\frac{\barc_{rr}}{\barr^2} \biggr),
\end{equation}
where $\barc_{ss}, \barc_{sr}$ and $\barc_{rr}$ are the elements of
$\barcov$.

The Jeffreys Bayes estimate for eigenratio (\ref{411}) was $\hate
=0.799$ (nearly the same as the MLE 0.793). Formula (\ref{62}) gave
$\hcv=0.002$, indicating that $\hate$ nearly equaled the exact Bayes
estimate $E\{t(\beta)|\hbet\}$. $B=10\mbox{,}000$ was definitely excessive.
Posterior parameters other than expectations are handled by other
well-known delta-method approximations. \textit{Note}: Discontinuous
parameters, such as the indicator of a parameter $\theta$ being less
than some value~$\theta_0$, tend to have higher values of~$\hcv$.

As far as external accuracy is concerned, the parametric bootstrap can
be employed to assess its own sampling error, a
``bootstrap-after-bootstrap'' technique in the terminology of \citet
{1992}. Suppose we have calculated some Bayes posterior estimate $\hatq
=Q(\hbet)$, for example, $\hate$ or a credible limit, and wonder
about its sampling standard error, that is, its frequentist
variability. As an answer, we sample $K$ more times from $f_{\hbet
}(\cdot)$,
%
\begin{equation}\label{63}
f_{\hbet}(\cdot)\longrightarrow\hgam_1,\hgam_2,
\ldots,\hgam_K,
\end{equation}
where the $\gamma$ notation emphasizes that these replications are
distinct from $\beta_1,\beta_2,\allowbreak\ldots,\beta_B$ in (\ref{36}), the
original replications used to compute $\hatq$. Letting $\hatq_k=Q(\hgam_k)$, the usual bootstrap estimate of standard error for
$\hatq$ is
%
\begin{equation}\label{64}
\hse (\hatq )= \Biggl[\sum_{k=1}^K (
\hatq_k-\hatq_\cdot )^2 /(K-1)
\Biggr]^{1/2},
\end{equation}
$\hatq_\cdot=\sum\hatq_k/K$. $K=200$ is usually plenty for
reasonable estimation of se$(\hatq)$; see Table 6.2 of \citet{1993}.

This recipe looks arduous since each $\hatq_k$ requires $B$ bootstrap
replications for its evaluation. Happily, a simple reweighting scheme
on the original $B$ replications finesses all that computation. Define
%
\begin{equation}\label{65}
W_{ki}=f_{\beta_i} (\hgam_k ) /f_{\beta_i} (
\hbet ).
\end{equation}

\begin{lemma}\label{lem61}
If $\hatq$ is a posterior expectation $\hate=\sum t_i\pi_iR_i/\sum \pi_iR_i$, then the importance sampling estimate of $\hatq_k$ is
%
\begin{equation}\label{66}
\hatq_k=\sum_{i=1}^Bt_i
\pi_iR_iW_{ki} \Big/\sum
_{i=1}^B\pi_iR_iW_{ki}
\end{equation}
for general quantities $\hatq$, reweighting $\beta_i$ proportionately
to $\pi_iR_iW_i$ gives $\hatq_k$.
\end{lemma}

The proof of Lemma~\ref{lem61} follows immediately from
%
\begin{equation}\label{67}
R_iW_{ki}=f_{\beta_i} (\hgam_k )
/f_{\hbet}(\beta_i),
\end{equation}
which is the correct importance sampling factor for converting an
$f_{\hbet}(\beta)$ sample into an $f_\beta(\hgam_k)$ likelihood.
\textit{Note}: Formula (\ref{66}) puts additional strain on our
importance sampling methodology and should be checked for internal
accuracy, as in~(\ref{62}).

Formula (\ref{66}) requires no new computations of $t(\beta), \pi
(\beta)$ or $R(\beta)$, and in exponential families the factor $W_{ki}$
is easily calculated:
%
\begin{equation}\label{68}
W_{ki}=e^{(\alpha_i-\halp)'(\hgam_k-\hbet)},
\end{equation}
where $\alpha_i$ is the canonical vector in (\ref{31}) corresponding
to $\beta_i$. This usually makes the computation for the
bootstrap-after-bootstrap standard error (\ref{64}) much less than
that needed originally for $\hatq$. [Formula (\ref{65}) is invariant
under smooth transformations of $\beta$, and so $W_{ki}$ can be
calculated directly in other coordinate systems as a ratio of densities.]

A striking use of (\ref{64}) appears in the last two columns of Table
\ref{table2}, Section~\ref{sec5}. Let $t_4(\beta_i)$ be the
indicator function of whether or not model 4 minimized AIC for the
$i$th bootstrap replication: $\hate\{t_4(\beta)|\hbet\}=0.36$
according to the \textit{Bayes \%} column. However, its
bootstrap-after-bootstrap standard error estimate was $\hse=0.20$,
with similarly enormous standard errors for the other model selection
probabilities. From a frequentist viewpoint, data-based model selection
will be a highly uncertain enterprise here.

Frequentist assessment of objective Bayes procedures has been advocated
in the literature, for example, in \citet{berger} and \citet{gelman},
but seems to be followed most often in the breach. The methodology here
can be useful for injecting a note of frequentist caution into Bayesian
data analysis based on priors of convenience.

If our original data set $\by$ consists of $n$ i.i.d. vectors $y_i$,
as in Table~\ref{table1}, we can jackknife instead of bootstrapping
the $\hgam_k$'s. Now $\hgam_k$ is $\hbet$ recomputed from the data
set $\by_{(i)}$ having $y_i$ removed for $k=1,2,\ldots,n$. Formulas
(\ref{65})--(\ref{68}) still hold, yielding
%
\begin{equation}\label{610}
\hse_{\jack}= \Biggl[\frac{n-1}{n}\sum_{k=1}^n
(\hatq_k-\hatq_\cdot )^2 \Biggr]^{1/2}.
\end{equation}
An advantage of jackknife resampling is that the $\hgam_k$ values lie
closer to $\hbet$, making $W_{ki}$ closer to 1 and putting less strain
on the importance sampling formula (\ref{66}).

\section{Summary}\label{sec7}

The main points made by the theory and examples of the preceding
sections are as follows:
\begin{itemize}
\item The parametric bootstrap distribution is a favorable starting
point for importance sampling computation of Bayes posterior
distributions (as in Figure~\ref{fig2}).
\item This computation is implemented by reweighting the bootstrap
replications rather than by drawing observations directly from the
posterior distribution as with MCMC [formulas (\ref{37}), (\ref{38})].
\item The necessary weights are easily computed in exponential families
for any prior, but are particularly simple starting from Jeffreys
invariant prior, in which case they depend only on the deviance
difference $\Delta(\beta)$ [(\ref{39})--(\ref{312}), (\ref{321}),
(\ref{47}), (\ref{55})].
\item The deviance difference depends asymptotically on the skewness of
the family, having a cubic normal form (\ref{329}).
\item In our examples, Jeffreys prior yielded posterior distributions
not much different than the unweighted bootstrap distribution. This may
be unsatisfactory for single parameters of interest in multiparameter
families (Figure~\ref{fig3}).
\item Better uninformative priors, such as the Welch--Peers family or
reference priors, are closely related to the frequentist BCa
reweighting formula [(\ref{217}), Figures~\ref{fig2} and~\ref{fig6}].
\item Because of the i.i.d. nature of bootstrap resampling, simple
formulas exist for the accuracy of posterior computations as a function
of the number $B$ of bootstrap replications. [Importance sampling
methods can be unstable, so internal accuracy calculations, as
suggested following (\ref{62}), are urged.] Even with excessive
choices of $B$, computation time was measured in seconds for our
examples (\ref{62}).
\item An efficient second-level bootstrap algorithm
(``bootstrap-after-bootstrap'') provides estimates for the frequentist
accuracy of Bayesian inferences [(\ref{63})--(\ref{66})].
\item This can be important in assessing inferences based on formulaic
priors, such as those of Jeffreys, rather than on genuine prior
experience (last column, Table~\ref{table2} of Section~\ref{sec5}).
\end{itemize}

\begin{appendix}\label{app}
\section*{Appendix}

\textit{Transformation of coordinates}:
Let $J(\beta)$ be the Jacobian of the transformation $\gamma=m(\beta)$,
that is, the absolute value of the determinant of the Hessian matrix
$(\partial\beta_i/\partial\gamma_j)$. Then $\tilf_\gamma(\hgam
)=f_\beta(\hbet)J(\hbet)$ gives
%
\begin{equation}\label{a1}
\tilxi(\gamma)=\frac{f_{\hbet}(\hbet) J(\hbet)}{f_\beta(\beta
)J(\beta)
}=\xi(\beta)\frac{J(\hbet)}{J(\beta)}
\end{equation}
in (\ref{42}), and
%
\begin{eqnarray}\label{a2}
\tilr(\gamma)&=&\frac{\tilf_\gamma (\hgam )}{\tilf_{\hgam}(\gamma)} =
\frac{f_\beta(\hbet)}{f_{\hbet}(\beta)}\frac{J(\hbet)}{J(\beta)
}=R(\beta)\frac{J(\hbet)}{J(\beta)}
\nonumber\\[-8pt]\\[-8pt]
&=&\xi(\beta)e^{\Delta(\beta)}\frac{J(\hbet)}{J(\beta)}=\tilxi (\gamma) e^{\tildel(\gamma)},
\nonumber
\end{eqnarray}
since $\tildel(\gamma)=\Delta(\beta)$ by the transformation invariance
of the deviance.

For any prior density $\pi(\beta)$ we have $\tilpi(\gamma)=\pi
(\beta)
J(\beta)$ and
%
\begin{eqnarray}
\label{a3} \tilpi(\gamma)\tilr(\gamma)&=&\pi(\beta)J(\beta)R(\beta)J(\hbet ) /J(
\beta)
\nonumber\\[-8pt]\\[-8pt]
&=&J(\hbet)\pi(\beta)R(\beta).\nonumber
\end{eqnarray}
$J(\hbet)$ acts as a constant in (\ref{a3}), showing that (\ref{44})
is identical to (\ref{37}). This also applies to Jeffreys prior,
$\tilpi^{\jeff}(\gamma)$, which by design is transformation
invariant, yielding
(\ref{45}).
\begin{pf*}{Proof of Theorem~\ref{thm34}}
In a one-parameter exponential family, (\ref{313}) and (\ref{314}) give
%
\begin{equation}\label{a4}
\psi(\alpha)-\psi (\halp )\doteq\hbet \,d\alpha+\hatv (d\alpha)^2/2+
\hatu(d\alpha)^3/6
\end{equation}
and
%
\begin{equation}\label{a5}
\beta-\hbet\doteq\hatv \,d\alpha+\hatu(d\alpha)^2/2,
\end{equation}
where $d\alpha=\alpha-\halp, \hatv=V(\halp)$, and $\hatu=U(\halp
)$. Expression (\ref{312}) for $\Delta$ can be written as
%
\begin{equation}\label{a6}
\Delta= (\beta-\hbet )\,d\alpha+2 \bigl[\hbet \,d\alpha - (\psi-\hat\psi ) \bigr].
\end{equation}
Applying (\ref{a4}) and (\ref{a5}) reduces (\ref{a6}) to
%
\begin{eqnarray}
\label{a7} \Delta&\doteq&\tfrac16\hatu(d\alpha)^3=\tfrac16\hgam \bigl[
\hatv^{1/2} (\alpha-\halp ) \bigr]^3
\nonumber\\[-8pt]\\[-8pt]
&\doteq&\tfrac16\hgam \bigl[\hatv^{-1/2} (\beta-\hbet )
\bigr]^3=\tfrac16\hgam Z^3\nonumber
\end{eqnarray}
with $\hgam=\hatu/\hatv^{3/2}$ the skewness, the last line following
from $Z\equiv\hatv^{-1/2}(\beta-\hbet)\doteq\hatv^{1/2}(\alpha
-\halp)$ (\ref{a5}). Standard exponential family theory shows that
$Z\to\n(0,1)$ under repeated sampling, verifying\vspace*{1pt} the theorem
[remembering that the asymptotics here are for $\beta\sim f_{\hbet
}(\cdot)$, with $\hbet$ fixed]. The skewness $\hgam$ is then
$O(n^{-1/2})$, making $\Delta$ of order $O_p(n^{-1/2})$. The first
missing term in the Taylor expansion (\ref{a7}) for $\Delta$ is $\hat
\delta Z^4/12$, $\hat\delta$ the kurtosis, and is of order
$O_p(n^{-1})$.\vadjust{\goodbreak}

The multiparameter version of Theorem~\ref{thm34} begins by
considering a one-parameter subfamily of (\ref{31}) now indexed by
$\alpha$ rather than $\beta$,
%
\begin{equation}\label{a8}
f_a^{(v)} (\hbet )=f_{\halp+av} (\hbet
)=e^{(\halp+av)'\hbet-\psi(\halp
+av)}f_0 (\hbet ),
\end{equation}
where $v$ is some fixed vector in $\real^p$; $a$ here is not connected
with that in (\ref{217}). The deviance difference within $f_a^{(v)}$ is
%
\begin{equation}\label{a9}
\Delta^{(v)}(a)=\Delta (\halp+av )
\end{equation}
since deviance is entirely determined by the two densities involved.

The exponential family terms (\ref{31}) for family $f_a^{(v)}(\cdot)$ are
%
\begin{eqnarray}
\label{a10} \alpha^{(v)}&=&a,\qquad \hbet^{(v)}=v'\hbet,\qquad
\beta^{(v)}=v'\beta,
\nonumber\\[-8pt]\\[-8pt]
\hatv^{(v)}&=&v'\hatv v\quad\mbox{and}\quad
\hatu^{(v)}=\sum_{j=1}^p
\sum_{k=1}^p\sum
_{l=1}^p\hatu_{jkl}v_jv_kv_l,
\nonumber
\end{eqnarray}
giving skewness $\hgam^{(v)}=\hatu^{(v)}/\hatv^{(v)3/2}$. Applying
the one-dimensional result gives
%
\begin{equation}\label{a11}
\Delta (\halp+av )\doteq\frac16\hgam^{(v)}{Z^{(v)3}}\qquad
\mbox{with }Z^{(v)}=\frac{v'(\beta-\hbet
)}{(v'\hatv v)^{1/2}}.
\end{equation}
Since $v$ can be any vector, (\ref{a11}) describes the asymptotic form
of $\Delta(\cdot)$ in the neighborhood of $\halp$.
\end{pf*}
\begin{pf*}{Proof of Theorem~\ref{thm41}}
For a single observation $y\sim\n_d(\mu,\nsig)$, let $f_1$ and $f_2$
represent its density under $(\mu_1,\nsig_1)$ and $(\mu_2,\nsig_2)$, respectively. Then
%
\begin{eqnarray}\label{a12}
2\log\frac{f_1(y)}{f_2(y)}&=&\log\frac{|\nsig_1|}{|\nsig_2|}+(y-\mu_2)'
\nsig_2(y-\mu_2)\nonumber\\[-8pt]\\[-8pt]
&&{}-(y-\mu_1)'
\nsig_1(y-\mu_1).\nonumber
\end{eqnarray}
But if $y\sim\n_d(\mu_1,\nsig_1)$,
%
\begin{eqnarray}\label{a13}
&&
E_{f_1} \bigl\{(y-\mu_2)'
\nsig_2^{-1}(y-\mu_2) \bigr\}\nonumber\\[-8pt]\\[-8pt]
&&\qquad=(
\mu_2-\mu_1)'\nsig_2^{-1}(
\mu_2-\mu_1)+\tr\nsig_1\nsig_2^{-1}\nonumber
\end{eqnarray}
while $E_{f_1}\{(y-\mu_1)'\nsig_1^{-1}(y-\mu_1)\}=d$. Taking the
$f_1$ expectation of (\ref{a12}) gives the deviance
%
\begin{eqnarray}\label{a14}\quad
&&
D \bigl((\mu_1,\nsig_1),(\mu_2,
\nsig_2) \bigr)\nonumber\\[-8pt]\\[-8pt]
&&\qquad=\log|\nsig_2|/|\nsig_1|+(
\mu_2-\mu_1)'\nsig_2^{-1}(
\mu_2-\mu_1)+\tr\nsig_1\nsig_2^{-1}-d\nonumber
\end{eqnarray}
for sample size $n=1$. The deviance difference for sample size $n$
%
\begin{equation}\label{a15}
\Delta=\frac{n}2 \bigl\{D \bigl((\mu,\nsig),(\hmu,\hnsig)\bigr) -D
\bigl((\hmu,\hnsig),(\mu,\nsig) \bigr) \bigr\}
\end{equation}
is then seen to equal (\ref{47}).\vadjust{\goodbreak}

The density of $(\hmu,\hnsig)$ from a $\n_p(\mu,\nsig)$ sample of size
$n$ is proportional to
%
\begin{equation}\label{a16}
\bigl\{|\nsig|^{-1/2}e^{-{n}(\hmu-\mu)'\nsig^{-1}(\hmu
-\mu)/2} \bigr\} \bigl\{|
\hnsig|^{({n-d-2})/2}e^{-{n}\tr
\nsig^{-1}\hnsig/2} /|\nsig|^{({n-1})/2} \bigr\}\hspace*{-28pt}
\end{equation}
yielding (\ref{46}).
\end{pf*}

\textit{The BCa weights}:
The BCa system of second-order accurate bootstrap confidence intervals
was introduced in \citet{1987} (Section 2 giving an overview of the
basic idea) and restated in weighting form (\ref{217}) in \citet
{1998}. The bias correction constant $z_0$ is obtained directly from
the MLE $\hthe$ and the bootstrap replication $\theta_1,\theta_2,\ldots,\theta_B$ according to
%
\begin{equation}\label{a17}
z_0=\Phi^{-1} \bigl(\# \{\theta_i\leq\hthe \} /B
\bigr\}.
\end{equation}
\citet{1992di} discuss ``ABC'' algorithms for computing $a$, the
acceleration constant. The program \texttt{abc2} is available in the
supplement to this article. It is very fast and accurate, but requires
individual programming for each exponential family. A more
computer-intensive R program, \texttt{accel}, which works directly
from the bootstrap replications $(\beta_i,t_i)$ [as in (\ref
{36}) and (\ref{37})], is also available in the supplement.

%
\begin{table}
\tablewidth=210pt
\caption{BCa constants $z_0$ and $a$ for our three examples}
\label{table3}
\begin{tabular*}{\tablewidth}{@{\extracolsep{\fill}}lccd{2.3}@{}}
\hline
&$\bolds{\hthe}$&$\bolds{z_0}$&\multicolumn{1}{c@{}}{$\bolds{a}$}\\
\hline
Student correlation&0.498&$-0.069$&0\\
Student eigenratio&0.793&$-0.222$&0\\
Prostate data Fdr(3)&0.192&$-0.047$&-0.026\\
\hline
\end{tabular*}
\end{table}

Table~\ref{table3} shows $z_0$ and $a$ for our three main examples.
Notice the especially large bias correction needed for the eigenratio.
\end{appendix}

\section*{Acknowledgment}

The author is grateful to Professor Wing H. Wong for several helpful
comments and suggestions.



\printaddresses

\end{document}